# Physics Educators as Designers of Simulation using Easy Java Simulation ( Ejs ) Part 2*

Loo Kang WEE

Ministry of Education, Education Technology Division, Singapore
wee_loo_kang@moe.gov.sg, weelookang@gmail.com

Abstract: To deepen do-it-yourself (DIY) technology in the physics classroom, we seek to highlight the Open Source Physics (OSP) community of educators that engage, enable and empower teachers as learners so that we create DIY technology tools-simulation for inquiry learning. We learn through Web 2 online collaborative means to develop simulations together with reputable physicists through the open source digital library. By examining the open source codes of the simulation through the Easy Java Simulation (EJS) toolkit, we are able make sense of the physics from the computational models created by practicing physicists. We will share newer (2010-present) simulations that we have remixed from existing library of simulations models into suitable learning environments for inquiry of physics. We hope other teachers would find these simulations useful and remix them that suit their own context and contribute back to benefit all humankind, becoming citizens for the world.

Abstract Footnotes: website prior to the meeting

http://www.phy.ntnu.edu.tw/ntnujava/index.php?board=28.0

*Extension of
Wee, L. K. (2010, 20 July). Physics Educators as Designers of Simulation using Easy Java Simulation (EJS). Paper presented at the American Association of Physics Teachers National Meeting Conference: 2010 Summer Meeting, Portland, Oregon, USA.

## I. INTRODUCTION

Easy Java Simulations (Ejs) is a software tool (java code generator) designed for the creation of discrete computer simulations.

I have created serveral computer models, also known as simulation, to allow our students to visualize Physics phenomena, using a free authoring toolkit called Easy Java Simulation [1].

Building on open source codes shared by the Open Source Physics (OSP) community and with help from Fu-Kwun's NTNUJAVA Virtual Physics Laboratory [2], I have customized serveral computer models as tools to support inquiry learning, downloadable from digital libraries in NTNUJAVA Virtual Physics Laboratory, creative commons attribution licensed.

Interested readers could refer to this Youtube [3] and blogpost [4] on the actual session.

## II. HOW CAN EJS BE USED?

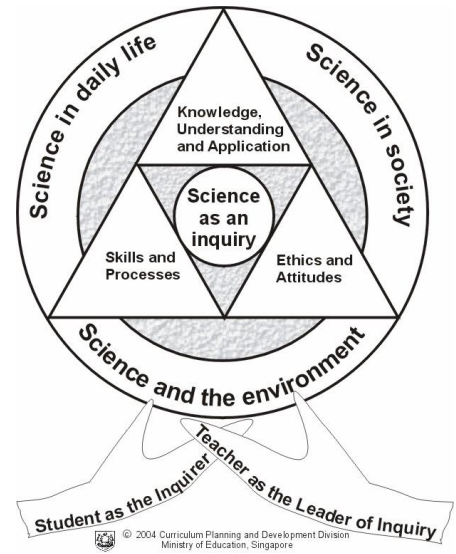

Figure 1.  The Ministry of Education Singapore Science Curriculum Framework [5]

My experiences in creating simulations for inquiry pedagogy enactment suggests students can conduct inquiry learning of science (Figure 1) while more ambitious tasks include creating and remixing their own simulations as representations of their understanding.

## III. FIVE COMPUTER MODELS

To add to the body of knowledge of teacher created-remixed simulations and align to the theme of the conference on the wave nature of light and matter, I would briefly highlight some of my design-ideas that I have made for the purpose of active inquiry learning on waves. Anyone is welcome to change the source codes of the simulations to suit their own needs, licensed creative commons attribution.

### A. Ripple Tank Model

This model (Figure 2) has 2 sources S1 and S2 and an investigative point P with top view as well as side view of the displacement that student usually have difficult visualizing. Intensity graph on the end of the top view are added thanks to another model [6] by Juan. A check-box for incoherence sources is also designed, an uncommon feature but available in our model for productive inquiry into conditions of incoherence. Lastly, a 3D view of the ripple tank seems to greatly enhance the visualization of the superposition of 2 point sources.

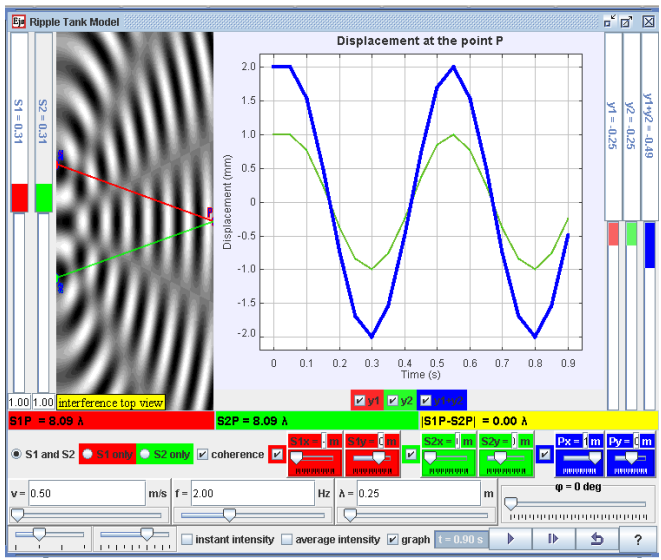

Figure 2. Ripple Tank Model [7] derived from Andrew's original work [8] with codes from Juan's work [6] customized to show top and side view of the ripple tank representations commonly taught in the A level Physics.

### B. 1 & 2 Slit(s) Diffraction Model

This model (Figure 3) allows the exploration of diffraction through 1 or 2 slits using the Huygens-Fresel principle. The 3D visualization is again useful with variables added to enable inquiry learning activities.

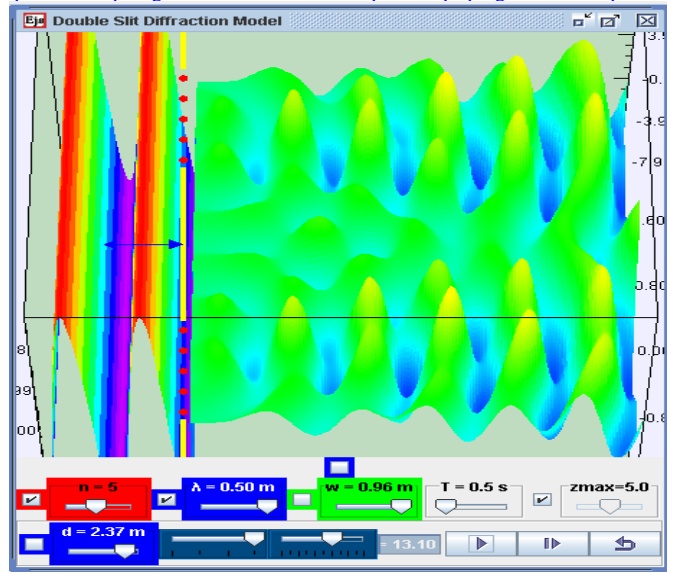

Figure 3. Double Slit Diffraction model [9] derived from Fu-Kwun's original work [10] showing 2 slits with 5 point sources in each slit with Huygens-Fresel principle, in a 3D view.

### C. Standing Wave in a Pipe Model

This model (Figure 4) is a remix customised to the Singapore syllabus with our teacher style of representation. It also has input fields to calculate according to the speed of sound using the formula $v = f \cdot \lambda$ so that student can verify their numercial answers to typical pen paper problems.

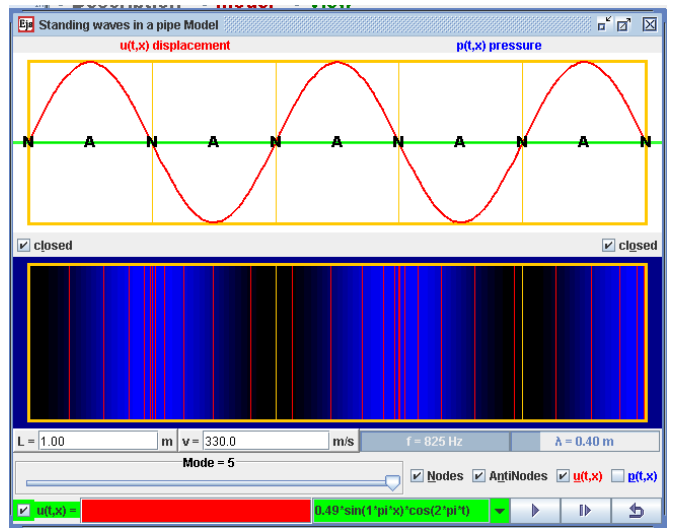

Figure 4. Standing Wave in a Pipe Model [11] derived from Juan's original work [12] showing side view and top view of the wave representation in a pipe with nodes and antinodes shown.

### D. Propagation EM Wave Model

This 3D model (Figure 5) is by Fu-Kwun Hwang which allows student to explore the horizontal circular motion provided only the radius is varied for perfect circular motion. My contributions include the forces visualization and variables that can be used for inquiry.

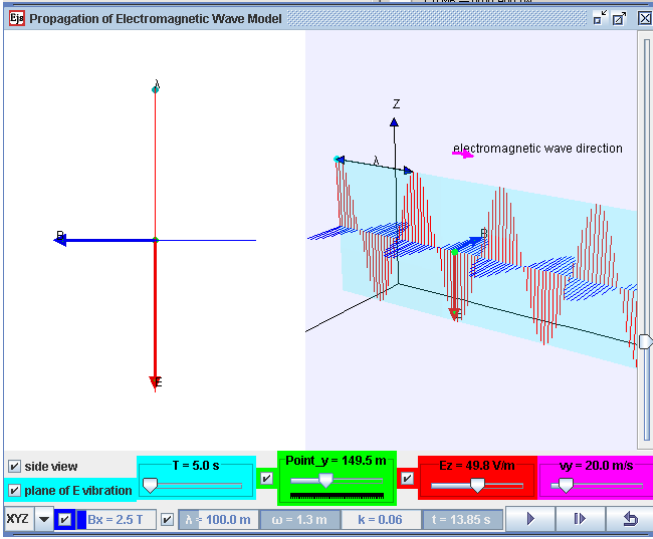

Figure 5. Propagation EM Wave Model [13] derived from Fu-Kwun's original work [14] with ideas from Juan's work [15] showing a 3D view of the electromagnetic wave property of nature of electric and magnetic wave.

*E. Blackbody Radiation Model*

This model (Figure 6) is very similar to PhET's interactive Flash program [16]. I realise how much I have learnt about computational physics as well as physics concepts modelled when I add new feature-ideas into this and other Ejs models.

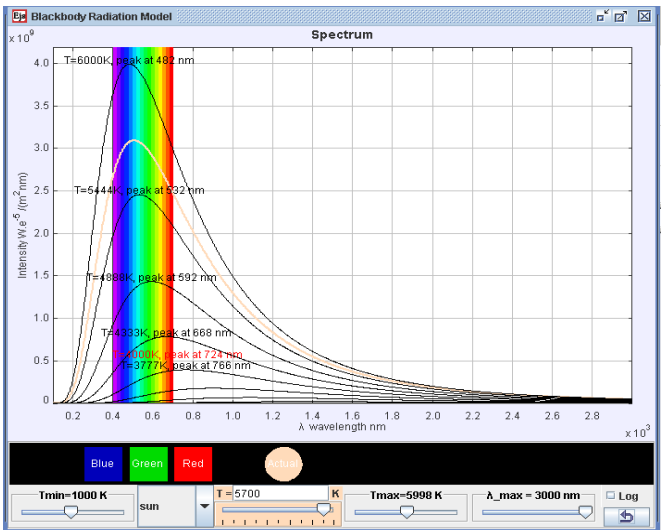

Figure 6. Blackbody Radiation Model [17] derived from Fu-Kwun's original work [18] with codes from Jose [19] that is added to visualise the real world observation.

## IV. CONCLUSION

I have shared some of my remixed simulations as an indication of how other teachers may deepen their professional practice in physics education, through informal professional learning with the Open Source Physics (OSP) and Easy Java Simulation (EJS) community.

I shared briefly some of the simulations that I have remixed from existing library of simulations models into suitable learning environments for inquiry of physics aligned with the theme of the conference, wave nature of light and matter.

I hope more teachers will find these simulations useful in their own classes and join in the journey of remixing simulations as a form of self directed professional learning to deepen our own teaching practices.

ACKNOWLEDGEMENT

I wish to acknowledge the passionate contributions of Francisco Esquembre, Fu-Kwun Hwang and Wolfgang Christian M. Belloni, A. Cox, W. Junkin, H. Gould, D. Brown, J. Tobochnik, Jose Sanchez, J. M. Aguirregabiria, S. Tuleja, M. Gallis, T. Timberlake, A. Duffy, T. Mzoughi, and many more in the EJS and OSP community for their simulations ideas and curriculum activities.

AUTHOR

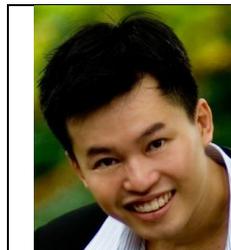

Loo Kang WEE is currently an educational technology specialist at the Ministry of Education, Singapore. He was a junior college physics lecturer and his research interest is in Open Source Physics tools like Easy Java Simulation for designing computer models and use of Tracker.